# Uncertainty Analysis of Melting and Resolidification of Gold Film Irradiated by Nano- to Femtosecond Lasers Using Stochastic Method

**Nazia Afrin**

**Yuwen Zhang**[1]
Fellow ASME

**J. K. Chen**
Fellow ASME

Department of Mechanical and Aerospace Engineering

University of Missouri

Columbia, MO 65211, USA

## ABSTRACT

A sample-based stochastic model is presented to investigate the effects of uncertainties of various input parameters, including laser fluence, laser pulse duration, thermal conductivity constants for electron, and electron-lattice coupling factor, on solid-liquid phase change of gold film under nano- to femtosecond laser irradiation. Rapid melting and resolidification of a free standing gold film subject to nano- to femtosecond laser are simulated using a two-temperature model incorporated with the interfacial tracking method. The interfacial velocity and temperature are obtained by solving the energy equation in terms of volumetric enthalpy for control volume. The convergence of variance (COV) is used to characterize the variability of the input parameters, and the interquartile range (IQR) is used to calculate the uncertainty of the output parameters. The IQR analysis shows that the laser fluence and the electron-lattice coupling factor have the strongest influences on the interfacial location, velocity, and temperatures.

## 1. Introduction:

At micro and nanoscales, ultra-fast laser material processing is a very important part in fabrication of some devices. Conventional theories established on the macroscopic level, such as heat diffusion assuming Fourier's law, are not applicable for the microscopic condition because they describe macroscopic behavior averaged over many grains [1]. For ultrashort laser pulses, the laser intensities can be high as $10^{12}$ W/m$^2$ or even higher up to $10^{21}$ W/m$^2$. During the laser interaction with materials, those electrons in the range of laser penetration of a metal material absorb the energy from the laser light and move with the velocity of ballistic motion. The hot electrons diffuse their thermal energy into the deeper part of the electron gas at a speed much slower than that of the ballistic motion. Due to the electron-lattice coupling, heat transfer to the lattice also occurs and a nonequilibrium thermal condition exists [2]. The nonequilibrium of electrons and the lattice are often described by two-temperature models by neglecting heat diffusion in the lattice [3, 4]. The accurate thermal response is only possible when the lattice conduction is taken into account in the physical model, particularly in the cases with phase change. Chen and Beraun [5] proposed a dual hyperbolic model which considered the heat conduction in the lattice.

---

[1] Corresponding author. Email: zhangyu@missouri.edu

In the physical process, melting in the lattice could take place for laser heating at high fluence. When the lattice is cooled, the liquid turns to solid via resolidification. The solid will be superheated in the melting stage, and the liquid will be undercooled in the resolidification stage. When the phase change occurs in a superheated solid or in an undercooled liquid, the solid-liquid interface can move at a very high velocity. Kuo and Qiu [6] investigated picoseconds laser melting of metal films using the dual-parabolic two-temperature model. Chowdhury and Xu [7] modeled melting and evaporation of gold film induced by a femtosecond laser. During the melting stage, the solid is superheated to above the normal melting temperature. During resolidification, the liquid is undercooled by conduction and the solid-liquid interface temperature can be below melting point. The solid-liquid interface can move at a high velocity which implies that the phase change is controlled by the nucleation dynamics, rather than energy balance [8].

In the melting and resolidification model of metal under pico- to femtosecond laser heating, the energy equation for electrons was solved using a semi-implicit scheme, while the energy and phase change equations for lattice were solved using an explicit enthalpy model [6,7]. The explicit scheme is easier to implement numerically than the implicit scheme for the enthalpy model [9]. Zhang and Chen proposed a fixed grid interfacial tracking method [10, 11] to solve rapid melting and resolidification during ultrafast short-pulse laser interaction with metal films. A nonlinear electron heat capacity obtained by Jiang and Tsai [12, 13] and a temperature-dependent coupling factor based on a phenomenological model [14] were employed in the two-temperature modeling [11]. The results showed that a strong electron-lattice coupling factor results in a higher lattice temperature which results a more rapid melting and longer duration of phase change.

Although the modeling of melting and resolidification of metal has significantly advanced in recent years, the inherent uncertainties of the input parameters can directly cause unstable characteristics of the output results. Among them, the laser fluence and pulse duration may fluctuate during the process. Moreover, the thermophysical properties of electrons and the lattice are not accurately determined at high temperatures. For example, the electron phonon coupling factor can either increase, decrease, or exhibit nonmonotonic changes with increasing electron temperature [15]. These parametric uncertainties may influence the characteristics of the phase change processes (melting and resolidification) which will affect the predictions of interfacial location, temperature and velocity and also the electron temperature. In the selective laser sintering (SLS), the fluence and width of laser pulses and the size of metal powder particles may influence the characteristics of the final product [16-20]. Therefore, study of parametric uncertainty is vital in simulation of the phase change of metal particles under nano- to femtosecond laser heating.

Sample-based stochastic model has been proposed to analyze the effects of the uncertainty of the parameters in order to integrate the parametric uncertainty distribution. Stochastic models possess some inherent randomness where the same set of parameter values and initial condition will lead to ensemble of different outputs. The stochastic model was applied on the non-isothermal filling process to investigate the effect of the uncertainty of parameters [21]. An improved simulation stochastic model was used in the ASPEN (a chemical process simulator) process simulator by Diwekar and Ruben [22]. The applications of the stochastic model in optical fiber drawing process [23, 24], thermosetting-matrix composite fabrication [25], microresonator [26] and proton change membrane (PEM) fuel cells [27] were found in the open literature. The sample–based stochastic model was applied to study the phase change of metal particle under uncertainty of particle size, laser properties and initial temperature to investigate the influences of the output parameters in the solid-liquid-vapor phase change of metal under nanosecond laser heating [28]. Convergence of variance (COV) was used to characterize the variability of the input parameters where the interquartile range (IQR) was used to measure the uncertainty of the output parameters.

In this paper, the sample-based stochastic model will be applied to study the melting and resolidification of gold film irradiated by nano to femtosecond laser under certain electron-phonon coupling factor, laser fluence, laser pulse width and constants for electron thermal conductivity to reveal the different influences of those parameters in the interfacial location, interfacial velocity, and interfacial and electron temperatures.

## 2. Physical model

A gold film with a thickness $L$ and an initial temperature $T_i$ is subjected to a laser pulse with a FWHM pulse width $t_p$ and fluence $J$ from the left hand surface. The energy equations of the free electrons and the lattice are:

$$C_e \frac{\partial T_e}{\partial t} = \frac{\partial}{\partial x}(k_e \frac{\partial T_e}{\partial x}) - G(T_e - T_l) + S' \tag{1}$$

$$C_l \frac{\partial T_l}{\partial t} = \frac{\partial}{\partial x}(k_l \frac{\partial T_l}{\partial x}) + G(T_e - T_l) \tag{2}$$

where $C$ represents heat capacity, $k$ is thermal conductivity, $G$ is electron-lattice coupling factor and $T$ is temperature. The heat capacity of electrons expressed as below is only valid for $T_e < 0.1 T_F$ with $T_F$ denoting Fermi temperature,

$$C_e = B_e T_e \tag{3}$$

where $B_e$ is a constant. According to Chen et al. [29], the electron heat capacity can be approximated by the following relationship:

$$C_e = \begin{cases} B_e T_e, & T_e < \frac{T_F}{\pi^2} \\ \frac{2B_e T_e}{3} + \frac{C'_e}{3}, & \frac{T_F}{\pi^2} \leq T_e < 3\frac{T_F}{\pi^2} \\ Nk_B + \frac{C'_e}{3}, & 3\frac{T_F}{\pi^2} \leq T_e < T_F \\ \frac{3Nk_B}{2} & T_e \geq T_F \end{cases} \tag{4}$$

where

$$C_e^{'} = B_e \frac{T_F}{\pi^2} + \frac{3\frac{Nk_B}{2} - \frac{B_e T_F}{\pi^2}}{T_F - \frac{T_F}{\pi^2}} (T_e - \frac{T_F}{\pi^2}) \tag{5}$$

The bulk thermal conductivity of metal at equilibrium can be represented as

$$k_{eq} = k_e + k_l \tag{6}$$

At the nonequilibrium condition the thermal conductivity of electrons depends on both electron and lattice temperatures. For a wide range of electron temperature ranging from room temperature, the thermal conductivity of electron can be measured as follows [30]:

$$k_e = \lambda \frac{(\vartheta_e^2 + 0.16)^{5/4}(\vartheta_e^2 + 0.44)\vartheta_e}{(\vartheta_e^2 + 0.092)^{1/2}(\vartheta_e^2 + \eta\vartheta_l)} \tag{7}$$

where $\vartheta_e = \frac{T_e}{T_F}$ and $\vartheta_l = \frac{T_l}{T_F}$ are dimensionless temperature parameters and $\lambda$ and $\eta$ are the two constants for the thermal conductivity of electrons. In general the values of those two constants for gold are $\lambda = 353$ W/mK and $\eta = 0.16$. For the low electron temperature limit ($\vartheta_e \ll 1$), the electron thermal conductivity can be expressed as

$$k_e = k_{eq}(\frac{T_e}{T_l}) \tag{8}$$

Under high energy laser heating, the electron and lattice temperatures change significantly which results in a temperature-dependent coupling factor in the ultra-fast laser heating. Chen et al. [14] proposed a relationship between electron and lattice temperatures for the coupling factor as follows:

$$G = G_{RT}[\frac{A_e}{B_l}(T_e + T_l) + 1] \tag{9}$$

where $A_e$ and $B_l$ are two material constants for the electron relaxation time; $G_{RT}$ is the room temperature coupling factor.

The heat source term in Eq. (1) can be represented as

$$S^{'} = 0.94 \frac{1-R}{t_p(\delta+\delta_b)[1-e^{-L/(\delta+\delta_b)}]} J \exp\left(\frac{x}{\delta+\delta_b} - 2.77(\frac{t}{t_p})^2\right) \tag{10}$$

where $R$ is reflectivity of the film, $J$ is the laser fluence, $\delta$ is the optical penetration depth, and $\delta_b$ is the ballistic range. At equilibrium, the bulk thermal conductivity of metal is measured as the summation of the electron thermal conductivity ($k_e$) and lattice thermal ($k_l$) conductivity. Free electrons are dominated in the heat conduction as the conduction mechanism is defined by the diffusion of free electron. So, for gold, the lattice and electron thermal conductivities are taken as 1% and 99% of the bulk thermal conductivity, respectively [31].

The energy balance at the solid-liquid interface in the system is given as

$$k_{l,s}\frac{\partial T_{l,s}}{\partial x} - k_{l,\ell}\frac{\partial T_{l,\ell}}{\partial x} = \rho_\ell h_m u_s \qquad (11)$$

where $\rho_\ell$ is the mass density of liquid, $h_m$ is the latent heat of fusion and $u_s$ is the solid-liquid interfacial velocity. For a metal under superheating the velocity of solid–liquid interface is expressed as follows [6]:

$$u_s = V_0[1 - \exp(-\frac{h_m}{R_g T_m}\frac{T_{l,I}-T_m}{T_{l,I}})] \qquad (12)$$

where $V_0$ is the maximum interfacial velocity, $T_{l,I}$ is the interfacial temperature and $R_g$ is the gas constant. The interfacial temperature could be higher than the normal melting temperature during melting and lower during solidification. The boundary conditions are given as

$$\left.\frac{\partial T_e}{\partial x}\right|_{x=0} = \left.\frac{\partial T_e}{\partial x}\right|_{x=L} = \left.\frac{\partial T_l}{\partial x}\right|_{x=0} = \left.\frac{\partial T_l}{\partial x}\right|_{x=L} = 0 \qquad (13)$$

The initial temperature conditions are

$$T_e(x, -2t_p) = T_l(x, -2t_p) = T_i \qquad (14)$$

The total computational domain is discretized with non-uniform grids. The implicit finite-difference equations are solved in each of the control volume (CV) and time step. The numerical solution starts from time -2$t_p$. During the solving process, the lattice temperature is set as interfacial temperature for that control volume that contains solid-liquid interface location. The energy equation in terms of enthalpy form is applied and solved for the solid liquid interface CV. The relationship of interfacial temperature and liquid fraction can be written by

$$C_{l,s}(T_{l,I})\frac{\partial T_{l,I}}{\partial t} + \rho_\ell h_m \frac{\partial f}{\partial t} = \frac{\partial}{\partial x}(k_l \frac{\partial T_l}{\partial x}) + G(T_e - T_l) \qquad (15)$$

where $T_{l,I}$ is the interfacial temperature, $C_{l,s}$ is the heat capacity at solid-liquid interface, and $f$ is the liquid fraction in the system. The liquid fraction is related to the location of the solid-liquid interface [11]. Before onset of melting, Eqs. (1) and (2) are solved simultaneously to obtain electron and lattice temperatures until the lattice temperature exceeds the melting point. Once it exceeds, the lattice temperature is set as the melting temperature and phase change will be considered in the system. After melting starts, an iterative procedure is applied to find the interfacial temperature and the interfacial location at each time step [10].

## 3. Stochastic modeling of uncertainty

Stochastic modeling is a process where the variability of the output parameters is evaluated based on the different combination of the input parameters [28]. In this paper, a sample-based stochastic model is used to study the melting and resolidification of the gold film under uncertain laser fluence, pulse width, coupling factor, and thermal conductivity of electrons to show the effects of the output parameters such as interfacial location, interfacial temperature, interfacial velocity and electron temperature. Figure 1 shows the detailed procedure of stochastic modeling.

In the stochastic modeling process, the first need is to quantify the degree to which the input parameters vary, and then to determine the appropriate number of combination of the input parameters to use with a stochastic convergence analysis. After determining the number of combinatiol of input parameters, one need to calculate the uncertainties of the input parameters though the deterministic physical model that was previously established. Eventually, the variability of the output parameters is quantified based on the uncertainty of input parameters. The coupling factor at room temperature between electron and lattice ($G_{RT}$), laser fluence ($J$), electron thermal conductivity constants ($\lambda$ and $\eta$), and laser pulse duration ($t_p$) are the input parameters whose uncertainties are going to be investigated. Due to the unavailability of experimental distribution of those uncertain parameters, it is acceptable to assume that all the input parameters follow Gaussian distributions of uncertainty [23]. The Gaussian distribution is defined by a mean value ($\mu$) and a standard deviation ($\sigma$), where the mean value is expressed by the nominal value of uncertainty parameters and the standard deviation represents the uncertainty of the input parameters. The coefficient of variance (COV) is an important parameter which represents the degree of uncertainty of the input parameters. The COV is defined as

$$COV = \frac{\sigma}{\mu} \tag{16}$$

After determining the distributions of the input parameters, a commonly used sampling method called Monte Carlo Sampling (MCS) is used to obtain the combination of the input parameters. According to the MCS input parameters are randomly selected from their prescribed Gaussian distributions and combined them together as one sample. Due to the high dependency on the number of the samples of input parameters on the variability of the output parameters, the exact number of samples of input parameters is determined carefully. In the stochastic convergence process, when the number of the sample increases the mean value and the standard deviation of input parameters converge to the nominal mean value and standard deviation of the Gaussian distribution. The mean value and standard deviation of the output parameters will also converge within a certain tolerance. After selecting the required number of samples for each input parameter, the physical model of melting and resolidification of gold film is solved. The effects of the input parameters variability on the output parameters uncertainty are evaluated by obtaining probability distribution of output parameters. The output parameters in this paper include interfacial location (s), interfacial temperature ($T_{l,I}$), interfacial velocity ($u_s$) and electron temperature ($T_e$). The probability distribution is calculated from the resulting set of the output parameters. The interquartile range (IQR) is a measurement of variability, based on dividing a data set into quartiles. It is defined as the difference between the 25$^{th}$ percentile and the 75$^{th}$ percentile,
$$IQR = P75 - P25 \tag{17}$$

## 4. Results and Discussions

The thermophysical and optical properties of pure gold film are [11]: $B_e$ = 70 J/m$^3$K, $A_e$ =1.2×10$^7$ K$^{-2}$s$^{-1}$ and $B_l$ =1.23×10$^{11}$ K$^{-1}$s$^{-1}$, $G_{RT}$ = 2.2×10$^{16}$ W/m$^3$K (solid) and 2.6×10$^{16}$ W/m$^3$K (liquid), $\rho$ =19.30×10$^3$ kg/m$^3$ (solid) and 17.28×10$^3$ kg/m$^3$ (liquid) reflectivity, $R$ = 0.6, $\delta$ = 20.6 nm, $\delta_b$ =105 nm, $T_m$ = 1336 K, $T_F$ = 6.42×10$^4$ K, $h_m$ = 6.373×10$^4$ J/kg, and $V_0$ =1300 m/s. The sample-based stochastic model provides the output parameter distributions with respect to the uncertain input parameter distributions. A large number of input samples is required to get the real distribution of the output parameters. Due to the difficulty in prohibitively intensive computation, it is important to find a minimum number of input samples ($N$) with which steady necessary output distributions can be generated.

To find the required number of $N$, we assume the nominal mean values of $G_{RT}$, $\lambda$, $\eta$, $J$ and $t_p$ are $2.2\times10^{16}$ W/m$^3$K, 353 W/mK, 0.16, 0.3 J/cm$^2$ and 20 ps, respectively. The coefficient of variance (COV) of each input parameter is set to be 0.02. Figure 2 represents the stochastic convergence analysis of the mean value of the input parameters $G_{RT}$, $\lambda$, $\eta$, $J$ and $t_p$. It is shown from this figure that when the number of samples is small, the mean values of the input parameters fluctuate significantly. For the value $N = 200$, the mean values of the input parameters oscillate in a smaller range, suggesting that a total of 200 samples should be sufficient for steady nominal mean values of input parameters. Figure 3 represents the stochastic convergence analysis of standard deviation of the five input parameters. It is shown that although the mean values of input parameters converges for 200 samples, the standard deviation still fluctuate. The reason behind this is that the deviation is a higher order moment which allows converging slower than the mean value. From Figure 3, it may conclude that the minimum number of the input samples is 300. After determining the minimum number of input samples, the stochastic convergence analysis for the mean value and standard deviation of the output parameters are obtained, as shown in Figures 4 and 5. It can be seen that when the number of the samples is beyond 300, the mean values of all the output parameters fluctuate in a smaller range (2.5%). Therefore, the minimum number of samples $N = 400$ is selected and used to calculate the results.

Figure 6 shows the typical distributions of the input parameters with the nominal mean values of $G_{RT}$, $\lambda$, $\eta$, $J$ and $t_p$ being $2.2\times10^{16}$ W/m$^3$K, 353 W/mK, 0.16, 0.3 J/cm$^2$ and 20ps respectively and the COV of each parameter being 0.02. Figure 7 gives the typical distribution of the output parameters s, $u_s$, $T_{l,I}$ and $T_e$. In the histograms, the distributions of the output parameters are no longer Gaussian due to the nonlinear effect in the solid liquid interface.

The IQRs of the output parameters s, $u_s$, $T_{l,I}$ and $T_e$ as functions of COV of the input parameters $G_{RT}$, $\lambda$, $\eta$, $J$ and $t_p$ are shown in the Fig. 8. When the COV of the one input parameter increases from 0.01 to 0.03 and the COVs of the other input parameters are kept constant at 0.01, the effect of that input parameter can be manifested. The IQRs of interfacial location, interfacial velocity, interfacial temperature and electron temperature significantly increases from 1.5 nm to 4.5 nm, 8.4 m/s to 23m/s, 298 K to 790 K, and 38 K to 110 K, respectively, when the COV of J increases from 0.01 to 0.03. The IQR of interfacial location, interfacial velocity, interfacial temperature and electron temperature significantly increases from 1.5 nm to 2.6 nm, 8.4 m/s to 16 m/s, 298 K to 550 K, and 38 K to 79 K, respectively, with the change of COV of $G_{RT}$ from 0.01 to 0.03. On the contrary, the COV of the thermal conductivity constants is relatively less impact to the interfacial location.

The IQR analysis of the interfacial velocity ($u_s$) shows that the laser influence $J$ is also most influential among the five input parameters. With the increment of COV of J from 0.01 to 0.03, the IQR of $u_s$ increases from 8.8 m/s to 23.1 m/s. As shown in Fig. 8, the order of influence of the COV of the five input parameters on the output parameters are $J$, $G_{RT}$, $t_p$, $\lambda$, and $\eta$. Figure 9 represents the IQRs of s, $u_s$, $T_{l,I}$ and $T_e$ for different laser influences with different COVs. As previously described, the COV of $J$ varies from 0.01 to 0.03 while the COVs of other parameters remain the same. It can be seen from Fig. 9 shows that for each laser influence the COV of $J$ significantly affects the IQR of the output parameters. With the increase of input value of COV from 0.01 to 0.03, the IQR of interfacial location, velocity, temperature and electron temperature increase from 1.575 nm to 5.25nm, 10m/s to 23 m/s, 398K to 790 K, and 44K to 115K, respectively for $J=0.4$ J/cm$^2$. That means, the larger the COV is, the more the IQR increases. Figure 10 represents the IQRs of s, $u_s$, $T_{l,I}$ and $T_e$ at different electron-lattice coupling factor ($G_{RT}$) with different COVs. Three values of $G_{RT}$, $2.1\times10^{16}$, $2.2\times10^{16}$ and $2.3\times10^{16}$ W/cm$^3$K, are considered with the COV ranging from 0.01 to 0.03, and the COVs of the other parameter remains the same. It is shown in Fig. 10 that for each $G_{RT}$, its COV significantly affects in the IQR of the output parameters. With the increase of input value of COV from 0.01 to 0.03, the IQR of interfacial location, interfacial velocity, temperature and electron temperature (K) increase from 1.58 nm to 2.85 nm, 8.5 m/s to 18 m/s, 305 K to 540 K, and 37 to79K, respectively, with GRT=$2.3\times10^{16}$ W/m$^3$K . The IQRs of s, $u_s$, $T_{l,I}$ and $T_e$ increase as the COV increases. The reason is that with the increase of the electron-phonon coupling factor, the hot electron heated up faster the metal lattice, leading to a more severe superheating process. Figures 9 and 10 indicate that the interfacial location, velocity and temperature and electron temperature greatly depends on the energy of laser and phonon-electron coupling factor.

## Conclusion

The sample-based stochastic model was applied to analysis the influence of parametric uncertainty on melting and resolidification of gold film subjected to nano- to femtosecond laser irradiation. This approach produces reasonable results with minimum number of combination of the input parameters to use with a stochastic convergence analysis. Rapid solid-liquid phase change was modeled using a two-temperature model with an interfacial tracking method. Temperature dependent electron heat capacity, thermal conductivity, and electron-lattice coupling factor were considered. The uncertainties of laser pulse fluence, pulse duration, electron-lattice coupling factor, and electron thermal conductivity on the results of solid-liquid interface temperature, interfacial velocity and location, and electron temperature were studied. The results show that the mean value and the standard deviation of laser influence and electron-lattice coupling factor have dominant effects on rapid phase change.

## Acknowledgement

Support for this work by the U.S. National Science Foundation under grant number CBET- 133611 is gratefully acknowledged.

## Nomenclature

$C_p$      heat capacity (J/kgK)

f      liquid factor

G      electron-lattice coupling factor (W/m$^3$K)

$G_{RT}$      electron-lattice coupling factor at room temperature (W/m$^3$K)

$h_m$      latent heat of fusion (J/kg)

| | |
|---|---|
| J | laser influence (J/cm$^2$) |
| K | thermal conductivity (W/mK) |
| P25 | 25$^{th}$ percentile |
| P75 | 75$^{th}$ percentile |
| $R_g$ | gas constant (J/kgK) |
| R | reflectivity |
| S' | laser source term |
| $T_{l,I}$ | interfacial lattice temperature (K) |
| $T_e$ | electron temperature (K) |
| $t_p$ | laser pulse (s) |
| $u_s$ | interfacial velocity (m/s) |
| $\delta$ | optical penetration depth (m) |
| $\delta_b$ | ballistic range (nm) |
| $\rho$ | density (kg/m$^3$) |
| $\lambda$ | thermal conductivity constant (W/mK) |
| $\eta$ | thermal conductivity constant |

Subscript

| | |
|---|---|
| e | electron |
| $\ell$ | liquid |
| l | lattice |
| m | melting |
| s | solid |

## References

[1] Tzou, D. Y., 1997, *Macro to Microscale Heat Transfer: The Lagging Behavior*, Taylor & Francis, Washington, DC.

[2] Hohlfeld, J., Wellershoff, S. -S., Güdde, J., Conrad, U., Jähnke, and V., Matthias, E., 2000, "Electron and Lattice Dynamics Following Optical Excitation of Metals," *Chemical Physics*, **251**(1-3), pp. 237-258.

[3] Anisimov, S. I., Kapeliovich, B. L., and Perel'man, T. L., 1974, "Electron Emission from Metal Surfaces Exposed to Ultrashort Laser," *Sov. Phys. Journal of Experimental and Theoretical Physics*, **39**(2), pp. 375-377.

[4] Qiu, T. Q., and Tien, C. L., 1993, "Heat Transfer Mechanism During Short-Pulse Laser Heating of Metals," *Journal of Heat Transfer*, **115**(4), pp. 835-841.

[5] Chen, J. K., and Beraun, J. E., 2001, "Numerical Study of Ultrashort Laser Pulse Interactions with Metal Films," *Numerical Heat Transfer*, A **40**(1), pp. 1-20.

[6] Kuo, L. S., and Qiu, T., 1996, "Microscale Energy Transfer During Picosecond Laser elting of Metal Films," ASME Natl. Heat Transfer Conf. **1**, pp. 149-157.

[7] Chowdhury, I. H., and Xu, X., 2003, "Heat Transfer in Femtosecond Laser Processing of Metal," *Numerical Heat Transfer,* A **44** (3), pp. 219 -232.

[8] Von Der Linde, D., Fabricius, N., Denielzik, B., and Bonkhofer, T., 1986, "Solid Phase Superheating During Picoseconds Laser Melting of Gallium Arsenide," *Materials Research Society Proceedings*, **74**, pp. 103-108.

[9] Voller, V.R., 1997, "An Overview of Numerical Methods for Solving Phase Change Problems," Advances in Numerical Heat Transfer, 1 ed. by W. J. Minkowycz, E. M. Sparrow (Taylor& Francis), Basingstoke, pp. 341-379.

[10] Zhang, Y., and Chen, J. K., 2008, "An Interfacial Tracing Method for Ultrashort Pulse Laser Melting and Resolidification of a Thin Metal Film," ASME *J. Heat Transfer*, **130**(6), 062401.

[11] Zhang, Y., and Chen, J. K., 2007, "Melting and Resolidification of Gold Film Irradiated by Nano-to Femtosecond Lasers," *Applied Physics* A, **88**, pp. 289-297.

[12] Jiang, L., and Tsai, H.-L., 2005, "Improved Two-Temperature Model and Its Application in Ultrashort Laser Heating of Metal Films," ASME *Journal of Heat Transfer*, **127**(10), pp. 1167-1173.

[13] Jiang, L., and Tsai, H.-L., 2005, "Energy Transport and Material Removal in Wide Bandgap Materials by a Femtosecond Laser Pulse," *International Journal of Heat and Mass Transfer*, **48**(3), pp. 487-499.

[14] Chen, J. K., Latham, W. P., and Beraun, J. E., 2005," The Role of Electron-Phonon Coupling in Ultrafast Laser Heating," *Journal of Laser Application*, **17**(1), pp. 63-68.

[15] Lin, Z., Zhigilei, L. V., and Celli. V., 2008, "Electron-Phonon Coupling and Electron Heat Capacity of Metals under Condition of Strong Electron-Phonon Equilibrium*," Physical Review* B, **77**, p. 075133.

[16] Beaman, J. J., Barlow, J. W., Bourell, D. L., Crawford, R.H., Marcus, H. L., and McAlea, K. P., 1997, *Solid Freedom Fabrication: A New Direction in Manufacturing*, Kluwer, Dordrecht, The Netherlands.

[17] Chen, T., and Zhang, Y., 2006, "Three-Dimensional Modeling of Selective Laser Sintering of Two-Component Metal Powder Layers," ASME *Journal of Manufacturing Science and Engineering*, **128**(1), pp. 299-306.

[18] Chen, T., and Zhang, Y., 2007, "Three-Dimensional Modeling of Laser Sintering of a Two-Component Metal Powder Layer on Top of Sintered Layers," ASME *Journal Manufacturing Science and Engineering*, **129** (3), pp. 575-582.

[19] Xiao, B., and Zhang, Y., 2008, "Numerical Simulation of Direct Metal Laser Sintering of Single-Component Powder on Top of Sintering Layers," ASME *Journal of Manufacturing Science*, **130** (4), p. 041002

[20] Fischer, P., Romano, V., Blatter, A., and Weber, H. P., 2005, "High Precision Pulsed Selective Laser Sintering of Metallic Powders," *Laser Physics Letters*, **2**(1), pp. 48-55.

[21] Padmanabhan, S. K., and Pitchumani, R., 1999, "Stochastic Modeling of Nanoisothermal Flow during Resin Transfer Molding Processes," *International Journal of Heat and Mass Transfer*, **42** (16), pp. 3057-3070.
[22] Diwekar, U.M., and Rubin, E. S., 1991, "Stochastic Modeling of Chemical Processes," *Computers Chemical Engineering*, **15** (2), pp. 105-114.
[23] Marwadi, A., and Pitchumani, R., 2008, "Numerical Simulations of an Optical Fiber Drawing Process under Uncertainty," Journal of Lighwave Technology, **26** (5), pp. 580-587.
[24] Myers, M. R., 1989, "A Model for Unsteady Analysis of Perform Drawing," AIChE J. **35** (4), pp. 592-602.
[25] Marwadi, A., and Pitchumani, R., 2004, "Cure Cycle Design for Thermosetting-Matrix Composites Fabrication under Uncertainty," *Annals Operations Research*, **132**(1), pp. 19-45.
[26] Mawardi, A., Pitchumani, R., 2005, "Design of Microresonators under Uncertainty," *Journal of Microelectromechanical Systems*, **14**(1), pp. 63-69.
[27] Marwadi, A., and Pitchumani, R., 2006, "Effect of Parameter Uncertainty on the Performance Variability of Proton Exchange Membrane (PEM) Fuel Cells," *J. Power Sources*, **160**(1), pp. 232-245.
[28] Pang, H., Zhang, Y., and Pai, P. F., 2013, "Uncertainty Analysis of Solid-Liquid-Vapor Phase Change of a Metal Particle Subject to Nanosecond Laser Heating," *Journal of Manufacturing Science and Engineering*, **135** (2), p. 021009.
[29] Chen, J. K., Beraun, J. E., and Tzou, D. Y., 2006, "A Semiclassical Two-Temperature Model for Ultrafast Laser Heating," *International Journal of Heat and Mass Transfer*, **49** (1-2), pp. 307-316.
[30] Anisimov, S. I., Rethfeld, B., 1997, "Theory of Ultrashort Laser Pulse Interaction with a Metal," Proc. SPIE, **3093**, 192-203.
[31] Klemens, P.G., And Williams, R.K., 1986, "Thermal Conductivity of Metals and Alloys," *International Materials Reviews*, **31**(1), pp. 197-215.

# Figure Captions

Figure 1    Sample-based stochastic model

Figure 2    Stochastic convergence analysis of mean value of the input parameters (a) $G_{RT}$, (b) $\lambda$, (c) $\eta$, (d) J and (e) $t_p$

Figure 3    Stochastic convergence analysis of standard deviation of the input parameters (a) $G_{RT}$, (b) $\lambda$, (c) $\eta$, (d) J and (e) $t_p$

Figure 4    Stochastic convergence analysis of mean value of the output parameters (a) s, (b) $u_s$, (c) $T_{l,I}$ and (d) $T_e$

Figure 5    Stochastic convergence analysis of standard deviation of the input parameters (a) s, (b) $u_s$, (c) $T_{l,I}$ and (d) $T_e$

Figure 6    Typical distributions of the input parameters (a) $G_{RT}$, (b) $\lambda$, (c) $\eta$, (d) J and (e) $t_p$

Figure 7    Typical distributions of the output parameters (a) s, (b) $u_s$, (c) $T_{l,I}$ and (d) $T_e$

Figure 8    The IQRs of the output parameters with different COVs of the input parameters (a) s, (b) $u_s$, (c) $T_{l,I}$ and (d) $T_e$

Figure 9    The IQRs of the output parameters with different values and COVs of $J$ (a) s, (b) $u_s$, (c) $T_{l,I}$ and (d) $T_e$

Figure 10    The IQRs of the output parameters with different values and COVs of $G_{RT}$ (a) s, (b) $u_s$, (c) $T_{l,I}$ and (d) $T_e$

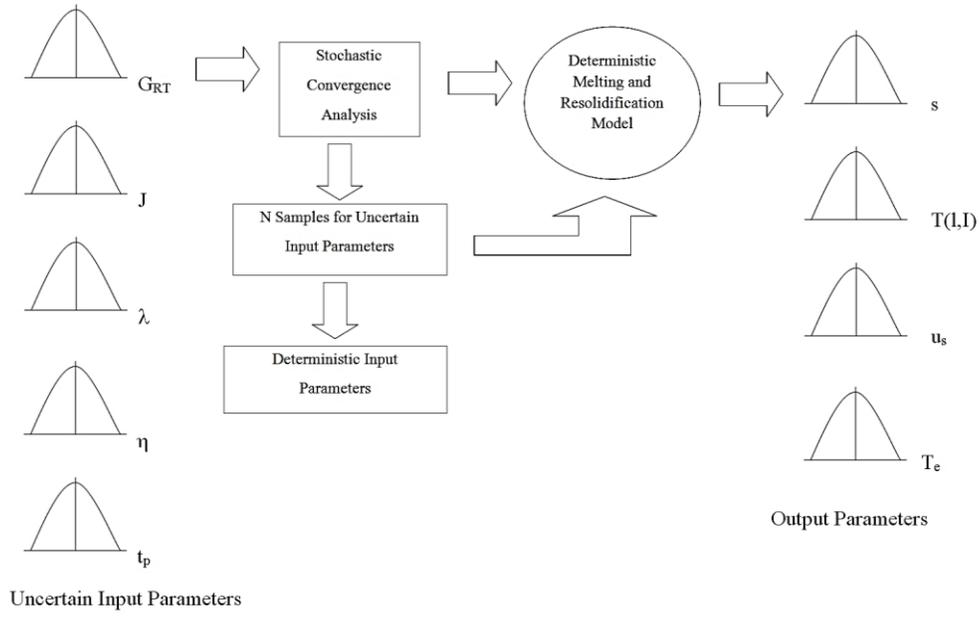

Figure 1

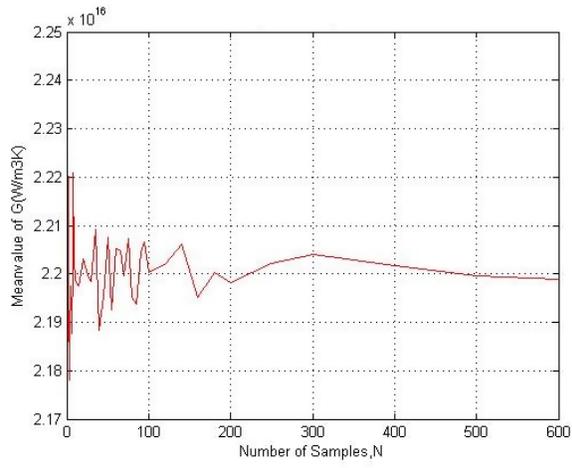

(a)

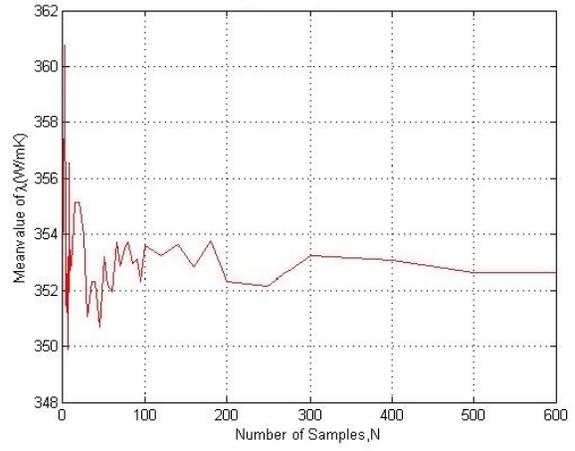

(b)

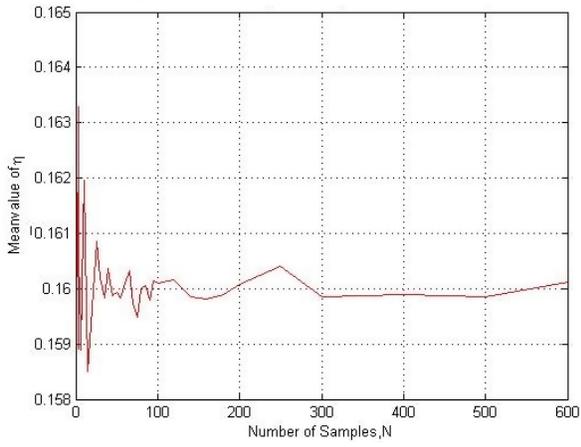

(c)

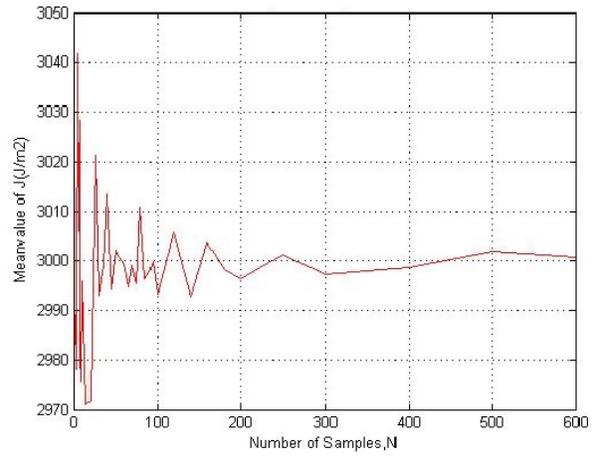

(d)

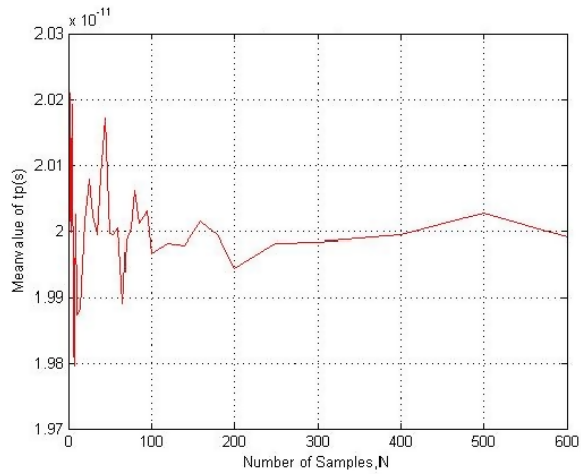

(e)

Figure 2

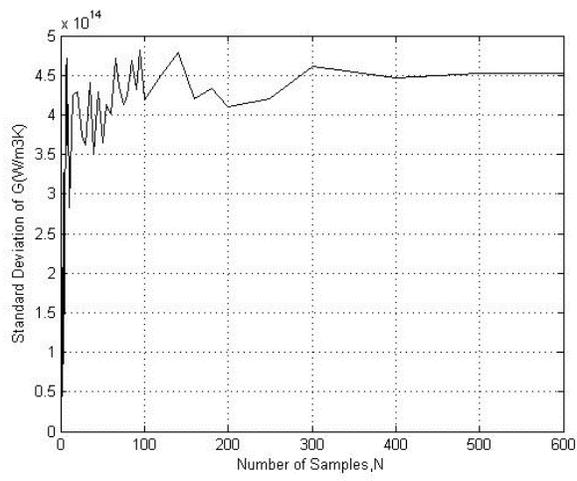
(a)

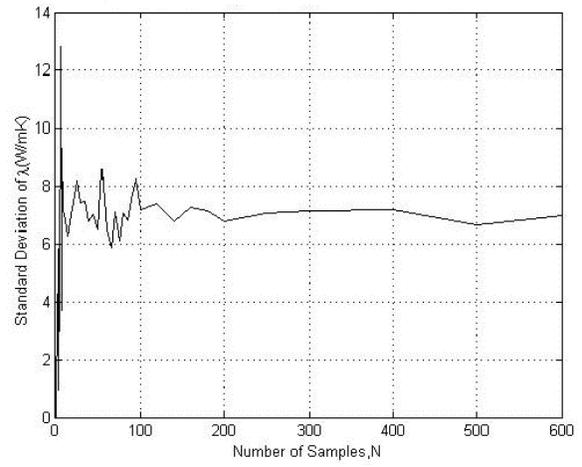
(b)

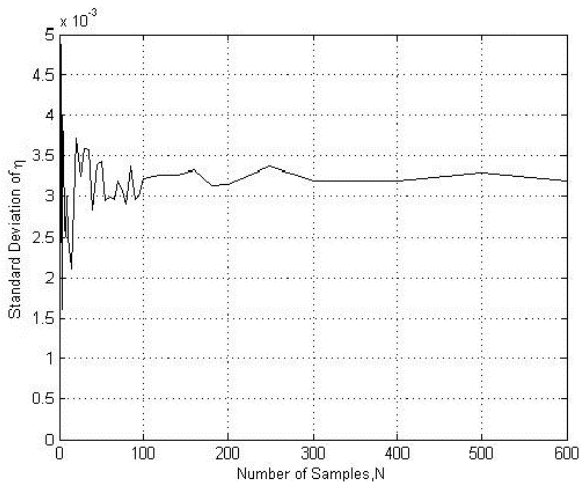
(c)

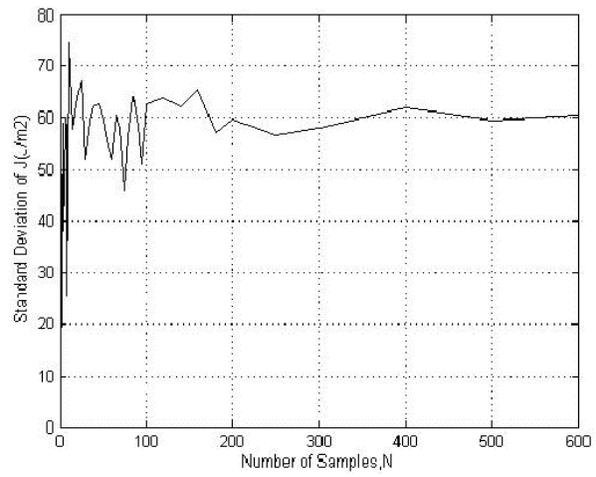
(d)

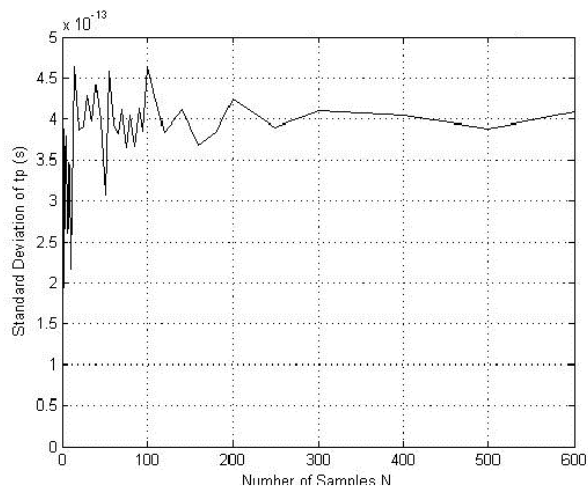
(e)

Figure 3

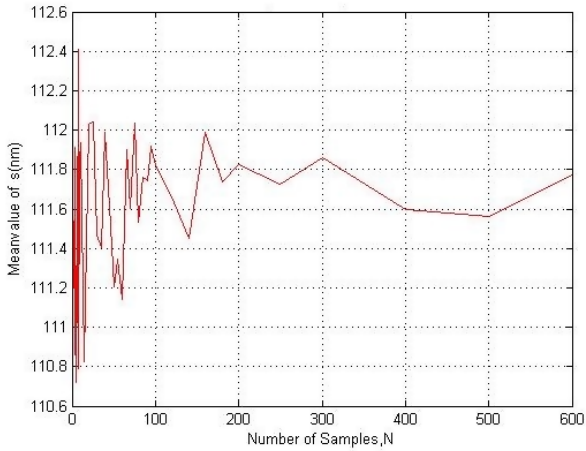
(a)

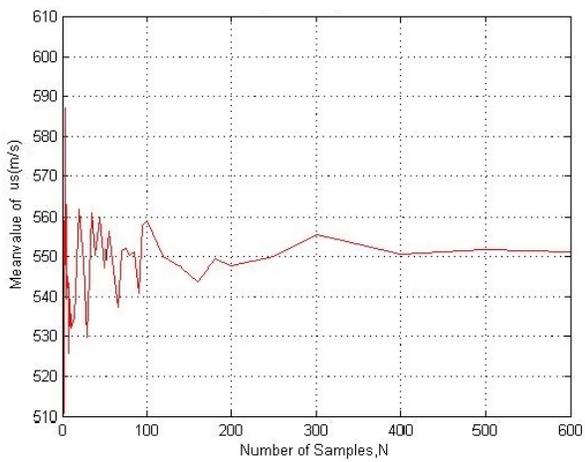
(b)

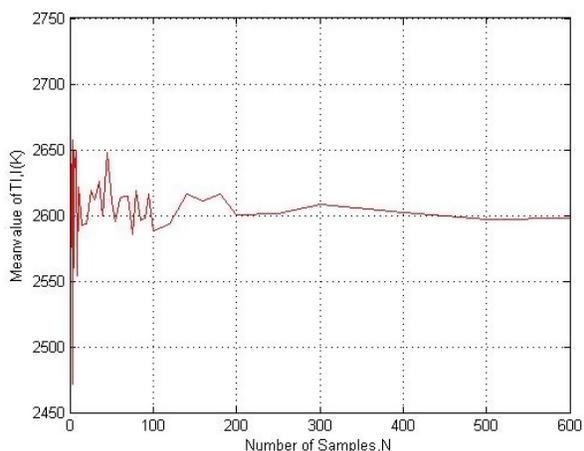
(c)

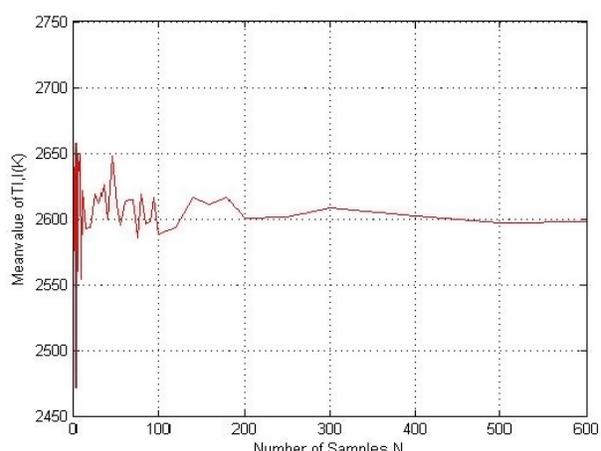
(d)

Figure 4

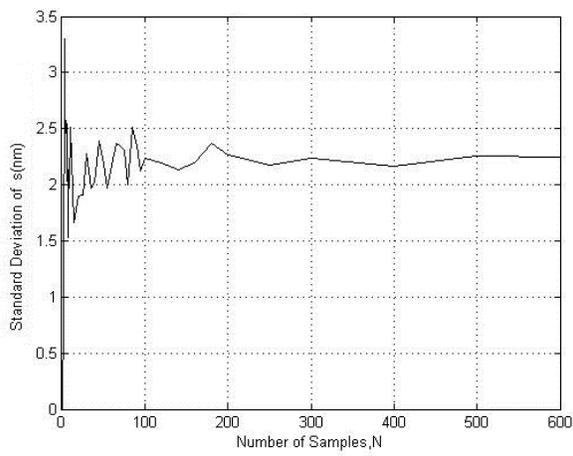
(a)

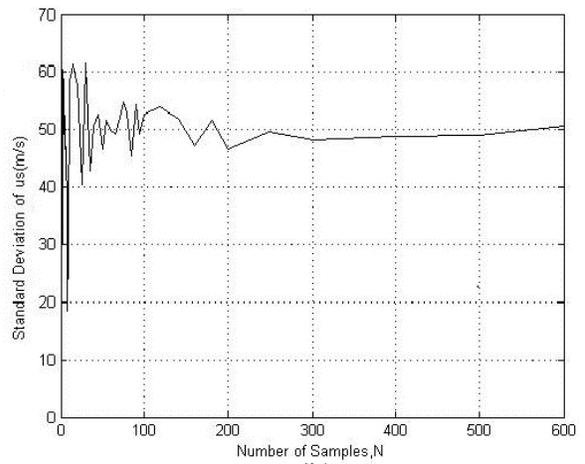
(b)

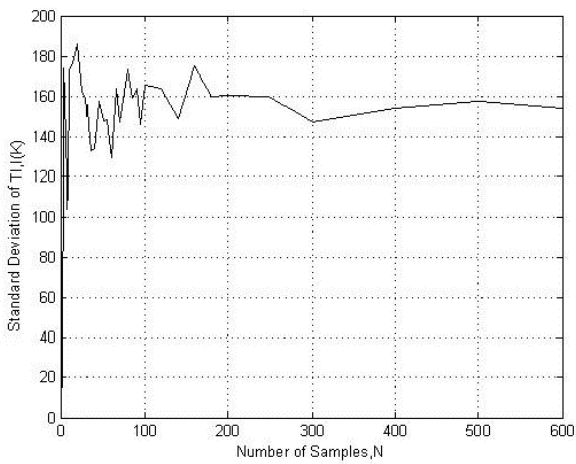
(c)

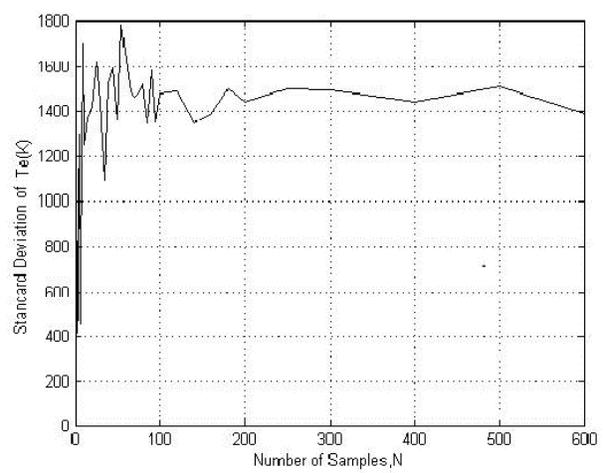
(d)

Figure 5

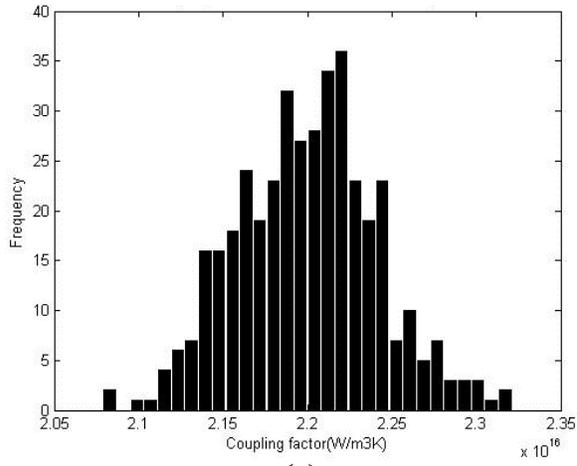

(a)

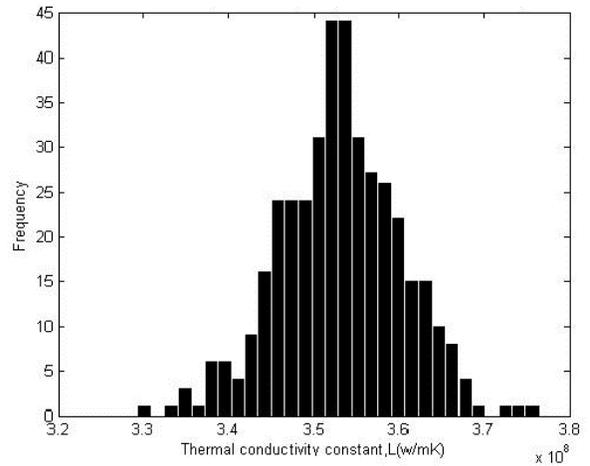

(b)

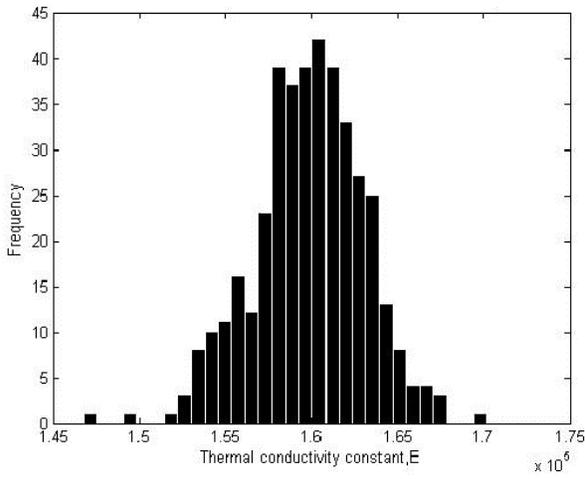

(c)

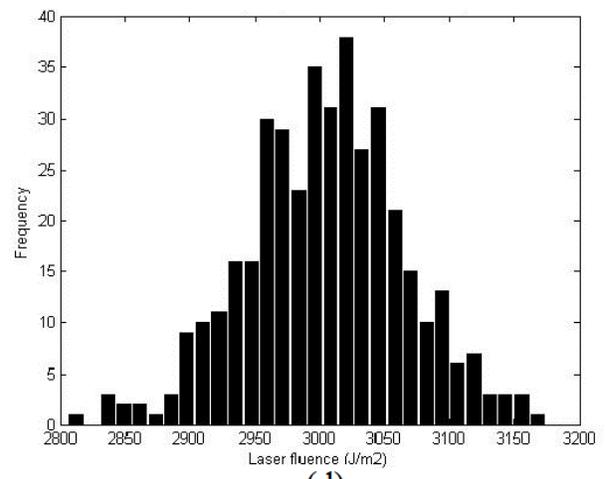

(d)

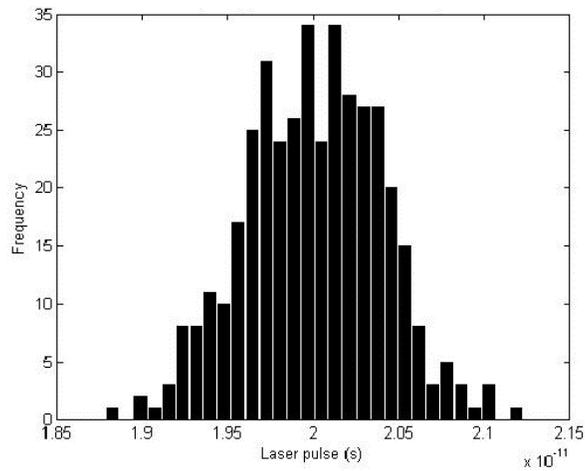

(e)

Figure 6

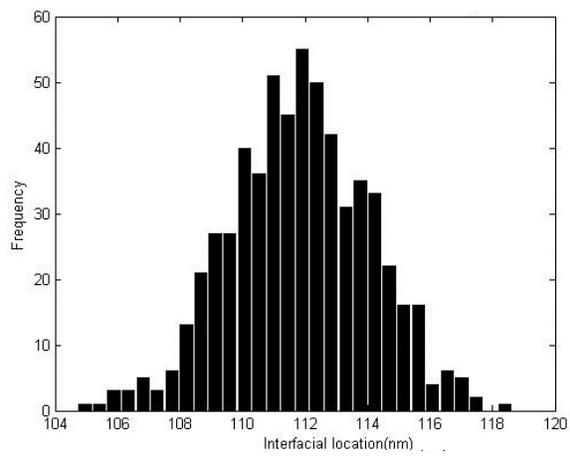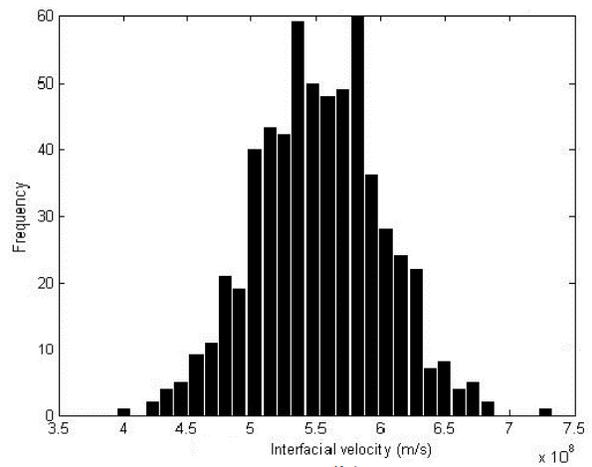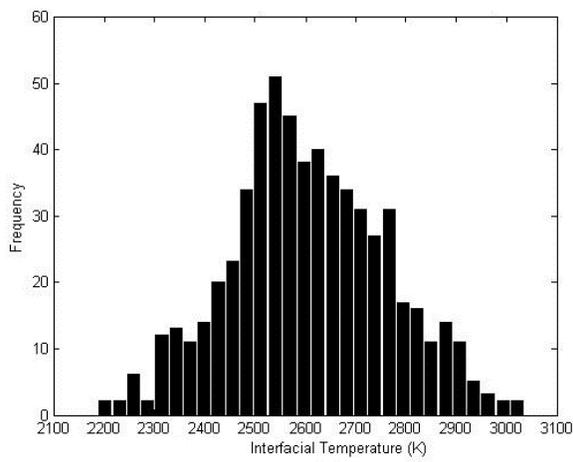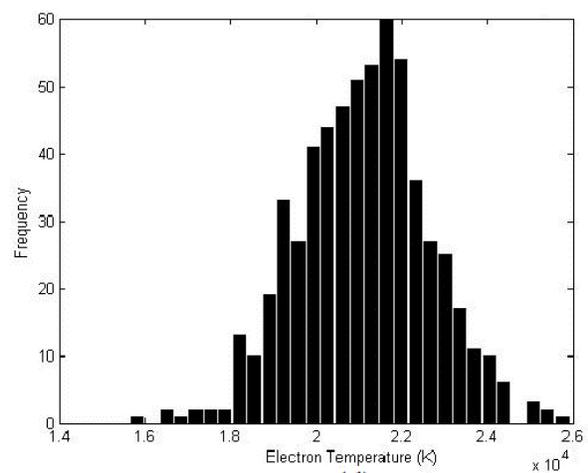

Figure 7

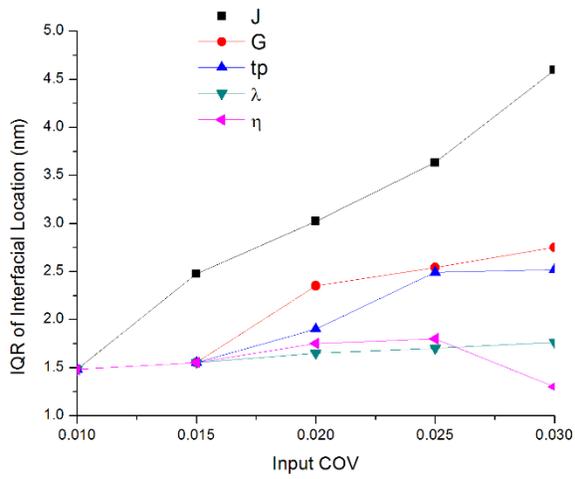 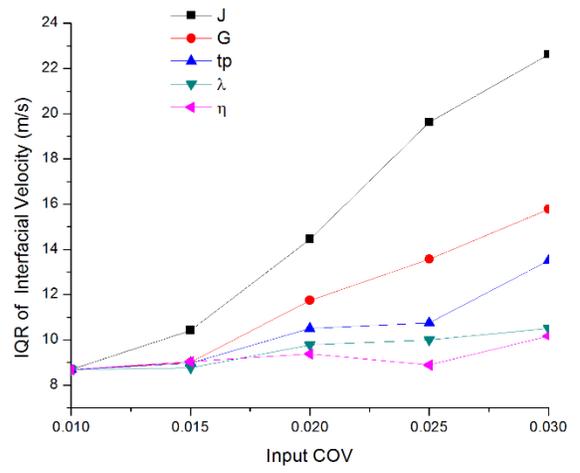
(a) (b)

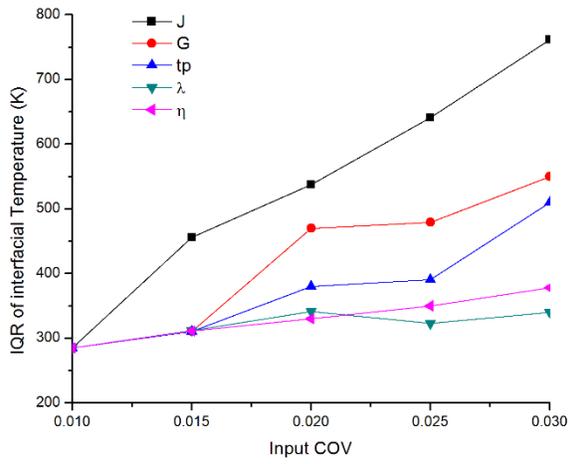 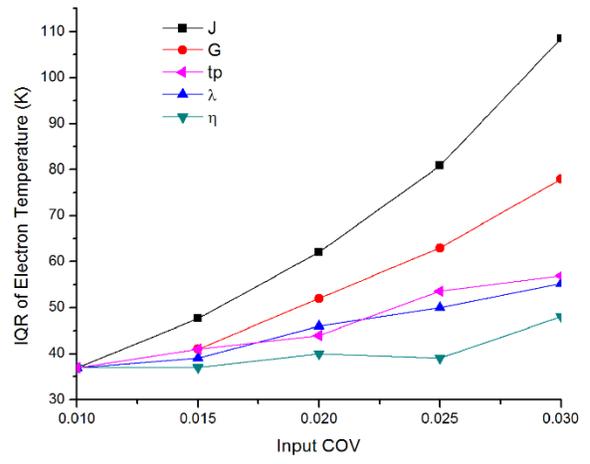
(c) (d)

Figure 8

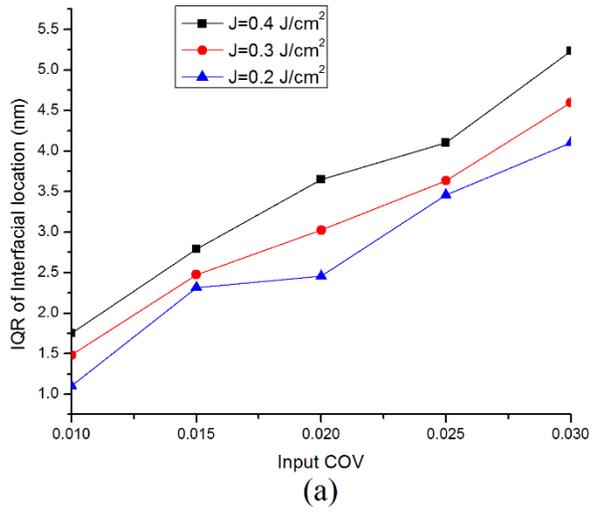
(a)

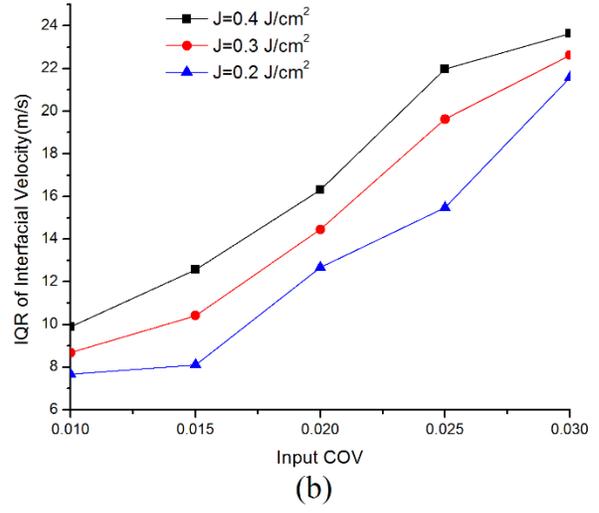
(b)

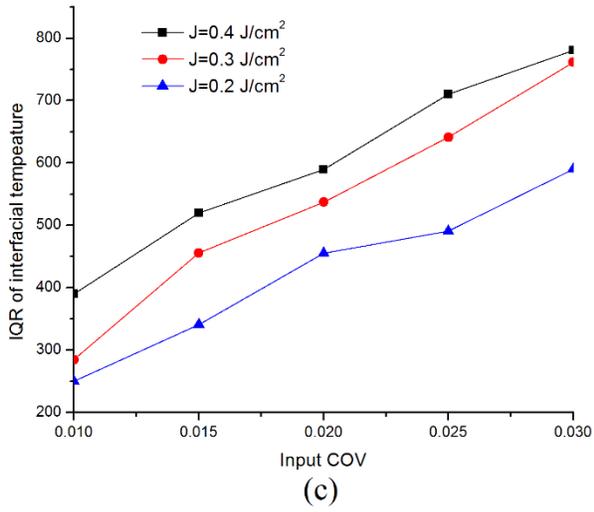
(c)

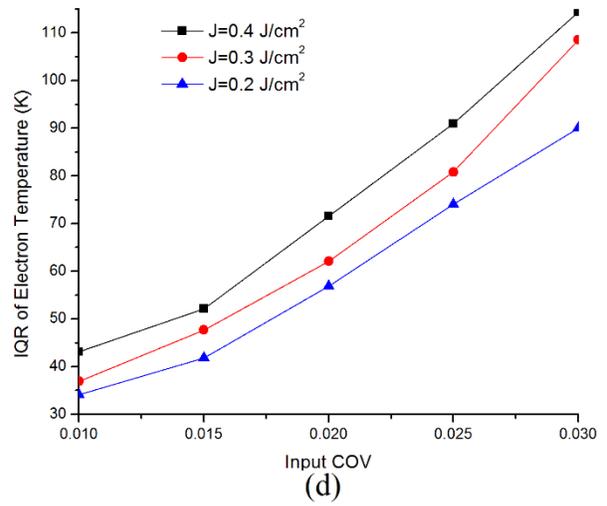
(d)

Figure 9

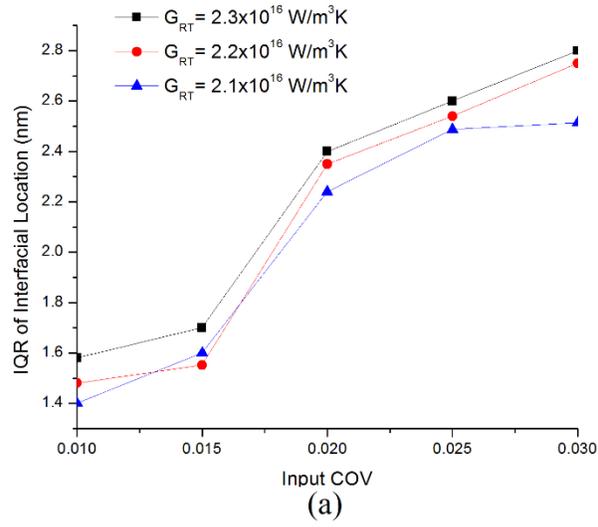
(a)

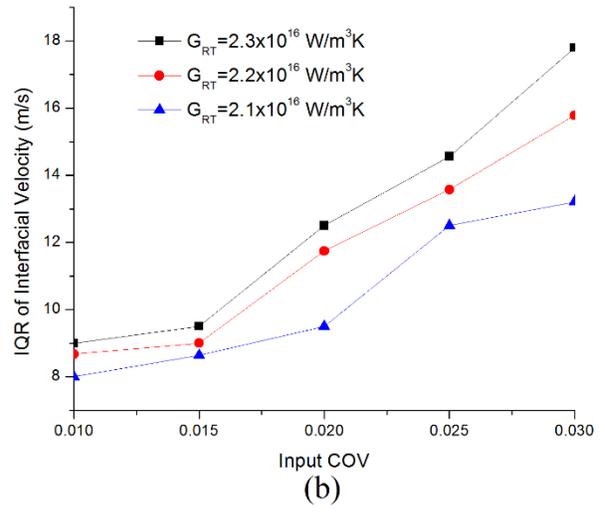
(b)

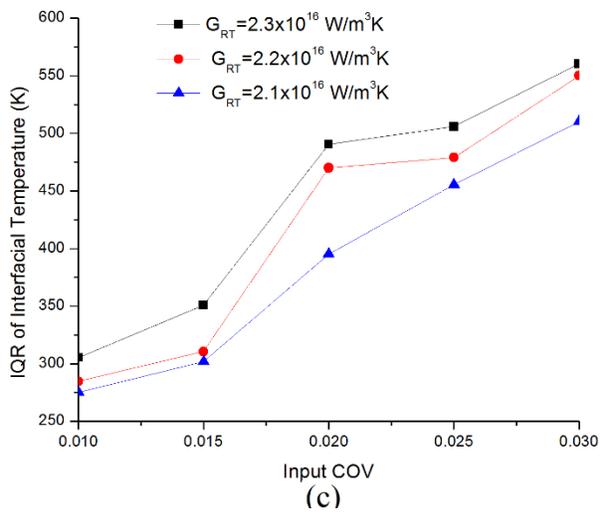
(c)

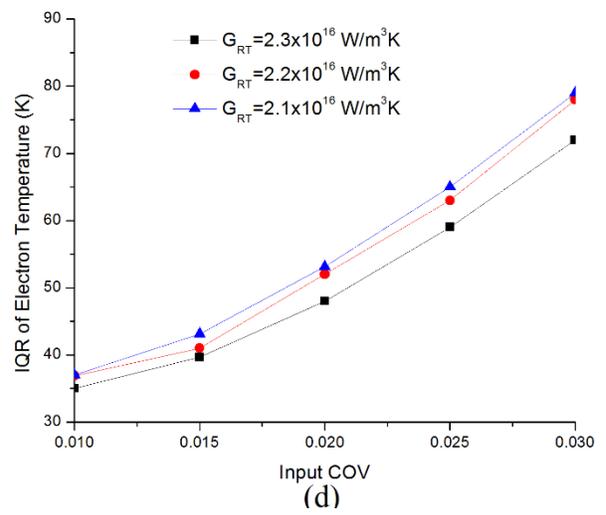
(d)

Figure 10